\documentclass[bm,aps,
amsfonts,amssymb,
preprint,
nofootinbib]{revtex4}

\usepackage[dvipdfmx]{graphicx}
\usepackage{natbib}
\usepackage{amsmath}

\usepackage{color}

\font\mybb=msbm10 at 12pt

\font\mybbsub=msbm10 at 8pt
\font\mybbsmall=msbm10 at 10pt

\font\myeu=eufm10 at 12pt

\font\myeusmall=eufm10 at 10pt

\def\bb#1{\hbox{\mybb#1}}

\def\bbsub#1{\hbox{\mybbsub#1}}
\def\bbsmall#1{\hbox{\mybbsmall#1}}

\def\frak#1{\hbox{\myeu#1}}

\def\fraksmall#1{\hbox{\myeusmall#1}}

\def\ZZ {\bb{Z}}
\def\ZZsub {\bbsub{Z}}
\def\ZZsmall {\bbsmall{Z}}

\def\QQ {\bb{Q}}

\def\QQsmall {\bbsmall{Q}}
\def\RR {\bb{R}}

\def\CCsub {\bbsub{C}}

\def\FF {\bb{F}}

\def\FFsmall {\bbsmall{F}}
\def\g {\frak{g}}

\def\gsmall {\fraksmall{g}}

\newcommand\beqa{\begin{eqnarray}}
\newcommand\eeqa{\end{eqnarray}}
\newcommand\n{\nonumber\\}

\begin{document}

{~}

\title{


Unifying error-correcting code/Narain CFT correspondences 
\\ via  
lattices over integers of cyclotomic fields 
}
\author{
Shun'ya Mizoguchi\footnote[1]{E-mail:mizoguch@post.kek.jp}
and Takumi Oikawa\footnote[2]{E-mail:oikawat@post.kek.jp}
}


\affiliation{\footnotemark[1]Theory Center, 
Institute of Particle and Nuclear Studies,
KEK\\Tsukuba, Ibaraki, 305-0801, Japan 
}

\affiliation{\footnotemark[1]\footnotemark[2]SOKENDAI (The Graduate University for Advanced Studies)\\
Tsukuba, Ibaraki, 305-0801, Japan 
}


\begin{abstract} 
We identify Narain conformal field theories (CFTs) that correspond to 
code lattices for quantum error-correcting codes (QECC) 
over integers of cyclotomic fields $\QQ(\zeta_p)$ $(\zeta_p=e^{\frac{2\pi i}p})$
for general prime $p\geq 3$. 
This code-lattice construction is a generalization of  
more familiar ones such as Construction A${}_{\CCsub}$ for ternary codes 
and (after the generalization stated below) Construction A for binary codes,
containing them as special cases. 
This code-lattice construction is redescribed in terms of root and weight lattices of
Lie algebras, which allows to construct lattices for codes  
over rings $\ZZ_q$ with non-prime $q$. 
Corresponding Narain CFTs are found for codes embedded into quotient rings 
of root and weight lattices of 
$ADE$ series, except $E_8$ and $D_k$ with $k$ even.
In a sense, this provides a unified description of the relationship between various 
QECCs over $\FF_p$ (or $\ZZ_q$) and Narain CFTs. 
%
A further extension on constructing the $E_8$ lattice from codes over the Mordell-Weil groups of extremal rational elliptic surfaces is also briefly discussed.
\end{abstract}

\preprint{KEK-TH-2660}
\date{October 16, 2024}

\maketitle

\newpage
\noindent
{\it Introduction.}~
An error-correcting code (ECC) is a technology for correcting errors 
that occur on a transmission path by adding redundancy to the transmitted information.
The relationship between classical error-correcting codes (CECCs) 
and chiral conformal field theories 
(CFTs) has been known for a long time, and has a long history of 
research \cite{Dolan1990, Dolan1996, Montague1994, Montague1993}. 
There was also groundbreaking progress in the construction of quantum error-correcting codes 
in the 1990s \cite{Shor9qubit,CSScode1,CSScode2,Gottesmanstabilizercode}. 
More recently, it has become clear that a certain class of quantum error-correcting codes 
(QECCs) can be linked to Narain CFTs \cite{Narain1986, Narain1987} 
through Lorentzian lattices \cite{Dymarsky2020}. 
An incomplete list of recent developments in this direction includes 
\cite{
Dymarsky:2020bps,Dymarsky:2020pzc,Dymarsky2021, Buican:2021uyp,
Yahagi2022, Furuta:2022ykh,Henriksson2023, Angelinos:2022umf,Henriksson:2022dml,
Dymarsky:2022kwb,Kawabata2023a, 
Furuta2023,Alam2023, Kawabata2023b, Kawabata2024a, 
Buican:2023bzl,Aharony:2023zit,Barbar:2023ncl,Buican:2023ehi,
Kawabata2024b}.

So far, in most (if not all) literature, the link between quantum ECCs and Narain CFTs
is made by first mapping a QECC belonging to a certain class (such as 
stabilizer codes \cite{Dymarsky2020
} 
or subsystem codes \cite{Kawabata2024b}) to a CECC of length $2n$, and then 
obtaining a code lattice using a procedure known as Construction A. 
On the other hand, connections between {\em classical} ECC and {\em chiral} CFT 
use a different kind of code-lattice constructions than Construction A, 
and exhibit remarkable properties. 
For example, it is known that the $E_8$ lattice, whose lattice theta function 
is a square root of the theta-function part of the partition function of the left-moving 16 bosons
of $E_8\times E_8$ heterotic string, can be constructed not only by Construction A 
from the extended Hamming code, but also by Construction ${\rm A}_{\CCsub}$ using 
Eisenstein integers from the so-called ``tetracode'', 
a ternary code of the simplest kind \cite{Conway2013}. Also it is known that the $E_8$ lattice can be 
derived from a length-two code over $\FF_5$ \cite{Ebeling} by using the lattice 
we will describe in this Letter below.
A natural question then arises: what kind of Narain CFTs correspond to codes over a 
general finite field $\FF_p$ if we exploit lattice constructions other than 
Construction A that are used in chiral CFTs?
This is the question we would like to address here.

In this Letter, we identify Narain CFTs that correspond to the code lattices 
over integers of the cyclotomic field $\QQ(\zeta_p)$ ($\zeta_p=e^{\frac{2\pi i}p}$) 
(See \cite{Conway2013,Ebeling} and references therein).
This lattice construction is a generalization of more familiar ones such as
Construction A for binary codes 
or Construction A${}_{\CCsub}$ 
for ternary codes 
\cite
{Conway2013}, 
and contains them as special cases \footnote{Although, 
strictly speaking, a cyclotomic field $\QQsmall(\zeta_p)$
is normally defined for an integer $p$ with $p\geq 3$, 
the obtained lattice over the algebraic integers has an isomorphism 
to the weight lattice of a Lie algebra, allowing a natural extension to $p=2$,
as we will explain below.
}. 
We will show that, for any prime $p$, the lattice so constructed from 
a length-$2n$ B-form code generated by the generator matrix
$\left(I~B\right)$,
where $I$ is the $n\times n$ identity matrix and $B$ is any $\FF_p$-valued 
$n\times n$ anti-symmetric matrix, coincides with 
the momentum-winding lattice of a Narain CFT compactified on a 
$(p-1)n$-dimensional torus with self-dual radius, 
where the metric $G_{\mu\nu}$ and the B-field $B_{\mu\nu}$ are given 
by the tensor products of $I$ and $C_p^{-1}$, and $B$ and $C_p^{-1}$, respectively, 
with $C_p^{-1}$ being the inner product matrix of the fundamental weights of 
$SU(p)$ \footnote{Up to the redundancy (\ref{redundancy}).  See below.}.

It turns out that the lattice constructed above essentially uses the isomorphism 
between $\ZZ_p$ and the quotient ring $\Lambda_W^{SU(p)}/\Lambda_R^{SU(p)}$, where 
$\Lambda_W^{SU(p)}$ and $\Lambda_R^{SU(p)}$ are the weight and root lattice of 
$SU(p)$, 
respectively.
Since this isomorphism itself is not limited to prime $p$ 
but also holds for general integers $p\geq 2$, 
we can define similar code lattices for non-prime integers $p\geq 2$ 
using the language of Lie algebras.
In this case, the corresponding codes are those over the ring $\ZZ_p$, 
and we can also find their Narain CFT counterparts.
This Lie algebraic characterization of the code lattices also allows generalizations 
to other $ADE$ Lie algebras, except $E_8$ and $D_k$ for even $k$ for reasons 
explained below.
%
This lattice construction is a special case of a more general one
using the lattice $\Lambda$ and its dual $\Lambda^*$
\cite{Dymarsky2020,Angelinos:2022umf,Aharony:2023zit,Barbar:2023ncl}, 
where we take as $\Lambda$ a root lattice of some Lie algebra.
Their explicit construction and their connection to the Narain CFTs have been
recently described in \cite{Angelinos'PhDthesis}, but the necessary B-field configuration
is not specified there.
%
%
\\
\\
\noindent
{\it 
Lattices over integers of a cyclotomic field.}
A {\it cyclotomic field} $\QQ(\zeta_p)$ is a number field obtained by adjoining 
a primitive $p$th root of unity $\zeta_p=e^{\frac{2\pi i}p}$ to the field of rational numbers $\QQ$.
Although it is defined itself for a general integer $p$ with $p\geq 3$, 
we first consider here the cases of odd prime $p$ for the purpose of the 
code-lattice construction \cite{Ebeling}. 
%
The definition is, as a set, 
\beqa
\QQ(\zeta_p)&:=&\left\{\left.
\sum_{i=0}^{p-2} a_i \zeta_p^i~\right|a_i\in\QQ~~\mbox{for}~~i=0,\ldots,p-2
\right\}.
\eeqa
This is a $(p-1)$-dimensional vector space over $\QQ$.

{\it Integers} of $\QQ(\zeta_p)$,
denoted by ${\frak O}$, is defined as
\beqa
{\frak O}&:=&\left\{\left.
\sum_{i=0}^{p-2} m_i \zeta_p^i~\right|m_i\in\ZZ~~\mbox{for}~~i=0,\ldots,p-2
\right\}.
\label{frakO}
\eeqa
Let $\alpha=\sum_{i=0}^{p-2} a_i \zeta_p^i$ be an element of $\QQ(\zeta_p)$.
Then the {\it trace} of $\alpha$, denoted by ${\rm Tr}\,\alpha$, is defined by
\beqa
{\rm Tr}\,\alpha&:=&(p-1)a_0-a_1-\cdots -a_{p-2}~~~\in\QQ.
\eeqa
Using this trace, we can define a symmetric bilinear form $\langle\ast,\ast\rangle$
on ${\frak O}$ as the function that sends $x,y\in{\frak O}$ to
\beqa
\langle x,y\rangle&:=&{\rm Tr}\frac{x\bar y}p.
\eeqa

Let us next consider the principal ideal ${\frak P}$ of ${\frak O}$ 
generated by $1-\zeta_p$. 
This means that ${\frak P}$ is a submodule of ${\frak O}$, consisting of 
polynomials in $\zeta_p$ with integer coefficients 
that are divisible by $1-\zeta_p$. 
One can show that ${\frak P}$ equipped with the bilinear form 
$\langle\ast,\ast\rangle$ is an even integral lattice isomorphic the $A_{p-1}$ root lattice.
Indeed, it is easy to verify that  the set of vectors 
\beqa
1-\zeta_p=:\vec\alpha_1,~~~ 
\zeta_p-\zeta_p^2=:\vec\alpha_2,~~~\ldots,~~~\zeta_p^{p-2}-\zeta_p^{p-1}=:\vec\alpha_{p-1}
\label{roots}
\eeqa
span ${\frak P}$, and the ``inner-product matrix'' $\langle \vec\alpha_i,\vec\alpha_j\rangle$
$(i,j=1,\ldots,p-1)$ is precisely the Cartan matrix of $A_{p-1}=SU(p)$.
Obviously from the factor theorem,  ${\frak P}$ is the kernel of the function 
$\rho: {\frak O}\rightarrow \FF_p$ 
in ${\frak O}$ given by
\beqa
\rho(\alpha)&:=&\sum_{i=0}^{p-2} a_i~~~\mbox{mod $p$}
\eeqa
for 
$\alpha=\sum_{i=0}^{p-2} a_i \zeta_p^i$.

Let ${\cal C}$ be a(n) (classical) error-correcting code over $\FF_p$ of 
length-$2n$ \footnote{ For later convenience, we set the code length to $2n$ here,
though the construction of the code lattice itself does not 
require that the code length be even.
}. 
The function $\rho$ is naturally extended to an $\FF_p^{2n}$-valued function on ${\frak O}^{2n}$, 
and we abuse the same notation.
The code lattice $\Gamma_{\cal C}$ is then defined as 
\beqa
\Gamma_{\cal C}&:=&\left\{
\rho^{-1}(c)\in {\frak O}^{2n}
\left|\,
c\in{\cal C}\subset \FF_p^{2n}
\right.
\right\}.
\label{Gamma_C}
\eeqa
Note that the pre-images of $\rho$ of $0,1,\ldots,p-1 \in \FF_p$ are 
$0,1,\ldots,p-1$ themselves as elements of ${\frak O}$ up to the kernel,
so a code $c \in \FF_p^{2n}$ can itself be thought of as a point 
${\frak O}^{2n}$ on the ``$2n$-fold real axes''.
Since the kernel of $\rho$ is ${\frak P}^{2n}$, $\Gamma_{\cal C}$
consists of elements of ${\frak O}^{2n}$ that are 
equal to any of the codes in ${\cal C}$
modulo ${\frak P}^{2n}$. 
%
%
%
\\
\\
\noindent
{\it ``Construction ${\rm A}_{\frak{g}}$'': A code-lattice construction in terms of Lie algebras.}
%
\\
As we saw above, $\frak P$ is isomorphic to the root lattice $\Lambda_R^{SU(p)}$ of $SU(p)$.
In fact, the ``integers'' $\frak O$ also allows a simple Lie-algebraic characterization: 
for any odd prime $p$, $\frak O$ (\ref{frakO})
endowed with the bilinear form 
$\langle\ast,\ast\rangle$ is isomorphic to the {\em weight} lattice 
$\Lambda_W^{SU(p)}$ of $SU(p)$.
Indeed, 
the $\ZZ$-span of vectors
\beqa
1=:\vec\omega_1,~~~ 
1+\zeta_p=:\vec\omega_2,~\ldots,
~~~1+\zeta_p+\cdots+\zeta_p^{p-2}=-\zeta_p^{p-1}=:\vec\omega_{p-1}
\eeqa
is obviously $\frak O$, and they satisfy
\beqa
\langle \vec\alpha_i,\vec\omega_j\rangle&=&\delta_{ij}
\eeqa
for $i,j=1,\ldots,p-1$.
These relations between $\frak P$, $\frak O$ and 
$\Lambda_R^{SU(p)}$, $\Lambda_W^{SU(p)}$ lead to an alternative definition 
of the code lattice to (\ref{Gamma_C}):
\beqa
\Gamma_{\cal C}&:=&\left\{
{\bf c}\,\vec{\omega}_1+\vec{\bf m}
\in (\Lambda_W^{SU(p)})^{2n}
\left|\,
{\bf c}\in{\cal C},~\vec{\bf m}\in (\Lambda_R^{SU(p)})^{2n}
\right.
\right\}.
\label{Gamma_CLambda_RLambda_W}
\eeqa
Here, we show the $2n$ component vectors in bold so that we can recognize that, 
for example, $\vec{\omega}_1$ and $\vec{\bf m}$ live in different spaces.
The code ${\bf c}$ is bolded accordingly.

Writing $\Gamma_{\cal C}$ as in (\ref{Gamma_CLambda_RLambda_W}), 
it is obvious that the lattice $\Lambda^{\CCsub}(\cal C)$ 
constructed from a ternary code ${\cal C}$ 
by Construction ${\rm A}_{\CCsub}$
is a special case of $\Gamma_{\cal C}$ when $p=3$.
Indeed, 
the definition of $\Lambda^{\CCsub}(\cal C)$ is
\beqa
\Lambda^{\CCsub}(\cal C)&:=&\left\{
\left.
\frac{c+(\omega-\bar\omega)m}{\sqrt{\tilde p}}
\right|\,c\in{\cal C},~~m\in{\cal E}^{2n}
\right\},
\eeqa
where ${\cal E}$ denotes the Eisenstein integers
\beqa
\cal E&:=&\{
m_1+m_2\,\omega\,|\,m_1,m_2\in\ZZ
\}.
\eeqa
$\omega$ is a cube root of unity $:=\frac{-1+\sqrt{3} i}2$. $c$ is a ternary code, i.e., 
a set of $2n$ symbols each of which can take on the values $0$, $1$, or $2$ mod $3$.
$\tilde p$ sets the overall scale of the lattice, so taking $\tilde p$ to be $3/2$, 
$\frac{(\omega-\bar\omega)m}{\sqrt{3/2}}=i\sqrt{2}m$ with $m\in{\cal E}^{2n}$ can be identified 
as the root lattice of $SU(3)$, in which the fundamental weight $\vec\omega_1$
is represented as $\frac1{\sqrt{3/2}}$.

Also, $\Gamma_{\cal C}$  (\ref{Gamma_CLambda_RLambda_W}) contains a special 
case of Construction A. If we take $p=2$,  $\Gamma_{\cal C}$  
(\ref{Gamma_CLambda_RLambda_W}) is nothing but the $k=2$ case of the lattice 
for binary codes
\beqa
\Lambda(\cal C)&:=&\left\{
\left.
\frac{c+k\,m}{\sqrt{k}}
\right|\,c\in{\cal C},~~m\in{\ZZ}^{2n}
\right\}
\eeqa
obtained by Construction A.
Note that Construction A constructs a lattice not only for binary codes but also 
for ones over $\FF_k$ for general prime $k$.
This number $k$ is the {\em level} of the lattice theta function that appears 
in the character formula of the level-$k$ $SU(2)$ affine Kac-Moody algebra.  
On the other hand, in $\Gamma_{\cal C}$  (\ref{Gamma_CLambda_RLambda_W}) 
the level is always fixed to $1$, but the code lattice over the same $\FF_k$ is realized by 
changing the {\em rank} of the underlying Lie algebra.  So in this sense, 
this might be seen as a code-lattice version of the level-rank duality.

As long as $p$ is an odd prime number, (\ref{Gamma_C}) 
and (\ref{Gamma_CLambda_RLambda_W}) give a completely equivalent definition.
On the other hand, (\ref{Gamma_C}) is initially defined only for odd prime $p$ 
as the vectors (\ref{roots}) would not be independent otherwise, 
but once we write the lattice $\Gamma_{\cal C}$ using the language of Lie algebras 
as in (\ref{Gamma_CLambda_RLambda_W}), we can relax this condition. 
Even if $p$ is not odd prime, $\Lambda_R^{SU(p)}$ is a sublattice of
$\Lambda_W^{SU(p)}$ of index $p$, whose quotient ring 
$\Lambda_W^{SU(p)}/\Lambda_R^{SU(p)}$ is $\ZZ_p=\ZZ/p\ZZ$ for general $p\geq 2$.
So if we take a code over the ring $\ZZ_p$ as ${\cal C}$ instead of one over $\FF_p$, 
 (\ref{Gamma_CLambda_RLambda_W}) is well-defined even for non-prime integers $p\geq 2$ 
 \footnote{If $p$ is prime, $\ZZsmall_p$ is promoted to $\FFsmall_p$. On the other hand, 
 if $\FFsmall_q$ is a finite field, it is known that $q$ must be written as $p^m$ 
 for some prime $p$ and some natural number $m$, but if $m\neq 1$, that is, if $q$ itself 
 is not a prime number, $\FFsmall_q\neq\ZZsmall_q$.}.

As we consider the construction of code lattices in this way, we can see that 
we can generalize these definitions a bit more. After all, the code lattice 
(\ref{Gamma_C}) or (\ref{Gamma_CLambda_RLambda_W}) is nothing but the image 
of the codes of the isomorphism from $\FF_p$ (or rather $\ZZ_p$) to the quotient 
ring $\Lambda_W^{SU(p)}/\Lambda_R^{SU(p)}$.
Thus it is also possible to replace $SU(p)$ by another simply-laced simple Lie algebra $\frak g$
and define a lattice $\Gamma_{\cal C}^{\frak g}$ 
associated with a code $\cal C$ over $\ZZ_q$ as \footnote{The idea of 
considering the quotient ring
$ \Lambda_W^{\fraksmall g}/\Lambda_R^{\fraksmall g} 
=\Lambda_R^{\fraksmall g*}/\Lambda_R^{\fraksmall g}$ itself 
(where $\Lambda_R^{\fraksmall g*}$ is the dual of the root lattice)  as a code 
is not new. What we want to do here is to ``construct" a lattice for a given code over 
$\FFsmall_p$ (or $\ZZsmall_q$) using the isomorphism 
and find a relationship between this lattice and the Narain CFT.}
\beqa
\Gamma_{\cal C}^{\frak g}&:=&\left\{
{\bf c}\,\vec{\omega}_{\rm gen}+\vec{\bf m}
\in (\Lambda_W^{\frak g})^{2n}
\left|\,
{\bf c}\in{\cal C},~\vec{\bf m}\in (\Lambda_R^{\frak g})^{2n}
\right.
\right\},
\label{Gamma_C^g}
\eeqa
where $\Lambda_R^{\frak g}$ and $\Lambda_W^{\frak g}$ are the root and weight lattice 
of $\frak g$, respectively, if $\Lambda_W^{\frak g}/\Lambda_R^{\frak g}=\ZZ_q$.
$\vec{\omega}_{\rm gen}$ is the generator of this $\ZZ_q$.
Such lattices have already been studied recently in \cite{Higa_talk}.

For example, if $\frak g=E_6$, $\Lambda_W^{E_6}/\Lambda_R^{E_6}=\ZZ_3$, 
so $\Gamma_{\cal C}^{E_6}$ is a lattice associated with a ternary code $\cal C$.
If $\frak g=E_7$, $\Lambda_W^{E_7}/\Lambda_R^{E_7}=\ZZ_2$, 
so $\Gamma_{\cal C}^{E_7}$ is a lattice for a binary code $\cal C$.
Note that we cannot take $E_8$ as $\frak g$ in (\ref{Gamma_C^g}) as 
the $E_8$ root lattice is self-dual $\Lambda_W^{E_8}=\Lambda_R^{E_8}$.
Finally, in the case of $D_k$, if $k$ is odd,
$\Lambda_W^{SO(2k)}/\Lambda_R^{SO(2k)}=\ZZ_4$
so we can treat it the same way, but if $k$ is even,
$\Lambda_W^{SO(2k)}/\Lambda_R^{SO(2k)}=\ZZ_2\times \ZZ_2$, 
so a different treatment is required.
For all these cases when $\Lambda_W^{\frak g}/\Lambda_R^{\frak g}=\ZZ_q$ and 
$\Gamma_{\cal C}^{\frak g}$ is well-defined, $\vec{\omega}_{\rm gen}$ can be
taken to be ``$\vec{\omega}_1$'', one of the fundamental weights of the nodes 
at the edges of the corresponding Dynkin diagram.

This code-lattice construction does not appear to have a special name. 
In this Letter, we call this lattice construction (\ref{Gamma_C^g}) 
``{\it Construction ${\rm A}_{\frak{g}}$}''.
Construction ${\rm A}_{\frak{g}}$ is equivalent to the code-lattice construction 
over integers of cyclotomic fields if $\frak{g}=SU(p)$ with odd prime $p$,
and is a natural generalization of the latter
\footnote{
In addition to $A_{p-1}$ for odd prime $p$ 
\cite{Ebeling}, 
the root lattices of $D_4$, $E_6$ and $E_8$ can also 
be embedded into the cyclotomic fields $\QQsmall(\zeta_8)$, $\QQsmall(\zeta_9)$ 
and $\QQsmall(\zeta_{20})$, respectively \cite{BayerFluckiger}, and 
that of $D_n$ with $n\geq 3$ can be embedded into $\QQsmall(\zeta_p)$ for prime $p$ 
such that $p\equiv 1$ mod $n$ \cite{AraujoJorge}. More generally, the root lattices of 
$D_n$ for odd $n$ and $A_n$ for $n=4m^2-1$ can also be embedded 
into a Galois extension $F/\QQsmall$ of odd degree $n$ \cite{HigaHirakawa}.
}.
We will see that the lattice so constructed 
allows a nice correspondence to  a Narain lattice of a certain class of CFT.
%
%
%
%
\\
\\
\noindent
{\it Code lattice\,/\,Narain CFT correspondence.}
%
In the remainder of this Letter, we identify the Narain CFT that corresponds to 
the code lattice (\ref{Gamma_CLambda_RLambda_W}) or  (\ref{Gamma_C^g}).
The general partition function of a two-dimensional scalar 
conformal field theory compactified on a $d$-dimensions torus 
is given by \cite{Narain1986,Narain1987}
\beqa
{\rm Tr}\,q^{L_0-\frac c{24}}\bar q^{\bar L_0-\frac c{24}}
&=&
\frac1{|\eta(\tau)|^{2c}}
\sum_{n_\mu\in\ZZsub^d}\sum_{w^\mu\in\ZZsub^d}
q^{
\frac1{4\alpha'}G^{\mu\nu}\big(
\alpha' \frac{n_\mu}R+(B_{\mu\rho}+G_{\mu\rho})w^\rho R
\big)
\big(
\alpha' \frac{n_\nu}R+(B_{\mu\sigma}+G_{\nu\sigma})w^\sigma R
\big)
}\n
&&~~~~~~~~\cdot
\bar q^{
\frac1{4\alpha'}G^{\mu\nu}\big(
\alpha' p_\mu+(B_{\mu\rho}-G_{\mu\rho})w^\rho R
\big)
\big(
\alpha' p_\nu+(B_{\mu\sigma}-G_{\nu\sigma})w^\sigma R
\big)
},
\label{partitionfunction}
\eeqa
where the central charge $c$ is $d$.
We first assume that the torus is a $(p-1)n$-dimensional torus defined by 
an $n$-fold direct product of the $SU(p)$ weight lattices 
whose metric is 
a tensor product of  $I$ 
and $C_p^{-1}$.
Explicitly,
\beqa
(G_{\mu\nu})_{\mu,\nu=1,\ldots,(p-1)n}&=&I \otimes C_p^{-1} ~=\,:~{\cal G}\n
&=&\frac1p \left(
\begin{array}{cccccc}
(p-1)I &(p-2)I &\cdots&\cdots&2I & I \\ 
(p-2)I &2(p-2)I & 2(p-3)I&\cdots&4I & 2I \\ 
\vdots&\vdots&\vdots&\vdots&\vdots&\vdots\\
2I & 4I &\cdots &2(p-3)I &2(p-2)I & (p-2)I \\
 I&2I &\cdots&\cdots&(p-2)I &  (p-1)I \\ 
\end{array}
\right),\n
\label{G}
\eeqa
where $\mu,\nu$ run over $1,2,\ldots,(p-1)n$.
This is the first step to embed a code into the quotient ring.
We also write the B-field 
$(B_{\mu\nu})_{\mu,\nu=1,\ldots,(p-1)n}$ as a 
$(p-1)n\times (p-1)n$ matrix ${\cal B}$, whose precise form will be 
determined shortly.
Note that the inverse of $C_p^{-1}$, that is $C_p$, is the inner-product 
matrix of the simple roots of $SU(p)$,
which is the Cartan matrix.

Using this ${\cal G}$ and ${\cal B}$ , it is easy to verify that 
the partition function (\ref{partitionfunction}) can be written as 
\beqa
{\rm Tr}\,q^{L_0-\frac c{24}}\bar q^{\bar L_0-\frac c{24}}
&=&
\frac1{|\eta(\tau)|^{2(p-1)n}}
\prod_{k=1}^{p-1}\left(
\sum_{{\bf n}_k\in\ZZsub^d}\sum_{{\bf w}_k\in\ZZsub^d}
\right)
q^{\frac{|\vec{\bf p}_L|^2}2}
\bar q^{\frac{|\vec{\bf p}_R|^2}2},
\eeqa
\beqa
\vec{\bf p}_L~=~\sum_{k=1}^{p-1}{\bf v}_k \vec{\alpha}_k, ~~~~~
\vec{\bf p}_R~=~\sum_{k=1}^{p-1}{\bf u}_k \vec{\alpha}_k,
\label{pLpR}
\eeqa
\beqa
\left(
\begin{array}{cccc}
{\bf v}_1 &{\bf v}_2 &\cdots &{\bf v}_{p-1} \\
\end{array}
\right)
&=&
\left(
\begin{array}{cccccccc}
\frac{{\bf n}_1}R &\frac{{\bf n}_2}R &\cdots &\frac{{\bf n}_{p-1}}R &
\mbox{\scriptsize ${\bf w}_1R$} &\mbox{\scriptsize ${\bf w}_2R$}  
&\cdots &\mbox{\scriptsize ${\bf w}_{p-1}R$}  \\
\end{array}
\right)
\left(
\begin{array}{c}
\mbox{\scriptsize $
\begin{array}{cccc}
I&&&\\
&I&&\\
&&\ddots&\\
&&&I
\end{array}$}
\\
\\
\frac12({\cal B}+{\cal G})^T\\
\\
\end{array}
\right),
\label{v}
\eeqa
\beqa
\left(
\begin{array}{cccc}
{\bf u}_1 &{\bf u}_2 &\cdots &{\bf u}_{p-1} \\
\end{array}
\right)
&=&
\left(
\begin{array}{cccccccc}
\frac{{\bf n}_1}R &\frac{{\bf n}_2}R &\cdots &\frac{{\bf n}_{p-1}}R &
\mbox{\scriptsize ${\bf w}_1R$} &\mbox{\scriptsize ${\bf w}_2R$}  
&\cdots &\mbox{\scriptsize ${\bf w}_{p-1}R$}  \\
\end{array}
\right)
\left(
\begin{array}{c}
\mbox{\scriptsize $
\begin{array}{cccc}
I&&&\\
&I&&\\
&&\ddots&\\
&&&I
\end{array}$}
\\
\\
\frac12({\cal B}-{\cal G})^T\\
\\
\end{array}
\right),
\label{u}
\eeqa
where we have set $\alpha'=2$.
The bold letters denote 
$n$-component vectors as before (but this time instead of $2n$-component ones).
$\vec{\alpha}_k$ $(k=1,\ldots,p-1)$ are the simple roots of $SU(p)$, so
\beqa
(\vec{\alpha}_k\cdot \vec{\alpha}_l)_{k,l=1,\ldots,p-1}
&=&C_p~=~
\left(
\begin{array}{ccccc}
2&-1&&&\\
-1&2&-1&&\\
&\ddots&\ddots&\ddots&\\
&&-1&2&-1\\
&&&-1 &2
\end{array}
\right),\\
(\vec{\omega}_k\cdot \vec{\omega}_l)_{k,l=1,\ldots,p-1}
&=&C_p^{-1}~=~
\frac1p \left(
\begin{array}{cccccc}
p-1 &p-2 &\cdots&\cdots&2 & 1\\ 
p-2 &2(p-2) & 2(p-3)&\cdots&4 & 2 \\ 
\vdots&\vdots&\vdots&\vdots&\vdots&\vdots\\
2 & 4 &\cdots &2(p-3) &2(p-2) & p-2\\
 1&2 &\cdots&\cdots&p-2 &  p-1 \\ 
\end{array}
\right),\n
\eeqa
where 
$\vec{\omega}_k$ $(k=1,\ldots,p-1)$ are the fundamental weights of $SU(p)$.

Thanks to the definition of $\cal G$ and the fact that 
\beqa
C_p^{-1}
\left(
\begin{array}{c}
\vec{\alpha}_1\\
\vdots\\
\vec{\alpha}_{p-1}
\end{array}
\right)
&=&
\left(
\begin{array}{c}
\vec{\omega}_1\\
\vdots\\
\vec{\omega}_{p-1}
\end{array}
\right),
\eeqa
we observe that 
\beqa
\vec{\bf p}_L-\vec{\bf p}_R
&=&
R\sum_{k=1}^{p-1}
{\bf w}_k \vec{\omega}_k.
\eeqa
Thus, taking 
$R$ to be the self-dual radius $\sqrt{2}$ $(=\sqrt{\alpha'})$, 
$\frac{\vec{\bf p}_L+\vec{\bf p}_R}{\sqrt{2}}$ runs over an $n$ copies
of the weight lattice $\Lambda_W^{SU(p)}$ of $SU(p)$.
%
%
%
%
%
%
On the other hand, the root lattice 
$\Lambda_R^{SU(p)}$ is a sub-lattice of $\Lambda_W^{SU(p)}$ 
of index $p$, and $k$ times the fundamental weight $k \vec\omega_1$ $(k=0,\ldots,p-1)$ 
can be taken to be the representatives of 
the $p$ classes of the quotient 
$\Lambda_W^{SU(p)}/\Lambda_R^{SU(p)}$. 
So we can write any point 
$\sum_{k=1}^{p-1}
{\bf w}_k \vec{\omega}_k$ 
on the $n$-fold tensor product of the weight lattices 
as
\beqa
\sum_{k=1}^{p-1}
{\bf w}_k \vec{\omega}_k
&=&
\sum_{k=1}^{p-1}
{\bf j}_k \vec{\alpha}_k\,+{\bf r}\vec\omega_1
\eeqa
for some ${\bf j}_k \in\ZZ^n$ and ${\bf r}\in\FF_p^n$ (or $\ZZ_p^n$ if $p$ is not prime).
Therefore,
\beqa
\frac{\vec{\bf p}_L-\vec{\bf p}_R}{\sqrt{2}}
&=&
\sum_{k=1}^{p-1}
{\bf j}_k \vec{\alpha}_k\,+{\bf r}\vec\omega_1\n
&\equiv&{\bf r}\,\vec\omega_1~~~\mbox{mod $(\Lambda_R^{SU(p)})^n$}.
\label{lefthalf}
\eeqa
We regard this ${\bf r}$ as the first length-$n$ part of a code over $\FF_p^n$ (or $\ZZ_p^n$)
arising from the generator matrix $\left(I~B\right)$.

We would like to look for a B-field such that the momentum lattice 
$\frac{\vec{\bf p}_L+\vec{\bf p}_R}{\sqrt{2}}$ yields the second half of the code. 
Since
\beqa
\vec{\bf p}_L+\vec{\bf p}_R
&=&
\sum_{k=1}^{p-1}
\left(\frac2R
{\bf n}_k\vec{\alpha}_k
+R\,
({\bf w} {\cal B})_k \,\vec{\alpha}_k
\right),
\eeqa
by setting $R=\sqrt{2}$, we have
\beqa
\frac{\vec{\bf p}_L+\vec{\bf p}_R}{\sqrt{2}}
&=&
\sum_{k=1}^{p-1}
({\bf n}_k+({\bf w}{\cal B})_k)\vec{\alpha}_k
\label{pL+pR/sqrt2}
\eeqa
If the rhs of (\ref{pL+pR/sqrt2}) becomes equal to 
${\bf r}B\,\vec\omega_1$ mod $(\Lambda_R^{SU(p)})^n$, 
it can be regarded as the right length-$n$ half of the length-$2n$ code 
whose left half is (\ref{lefthalf}).
For this to happen, we take 
\beqa
(B_{\mu\nu})_{\mu,\nu=1,\ldots,(p-1)n}&=&B \otimes C_p^{-1}~=\,:{\cal B}\n
&=&\frac1p \left(
\begin{array}{cccccc}
(p-1)B &(p-2)B &\cdots&\cdots&2B & B \\ 
(p-2)B &2(p-2)B & 2(p-3)B&\cdots&4B & 2B \\ 
\vdots&\vdots&\vdots&\vdots&\vdots&\vdots\\
2B & 4B &\cdots &2(p-3)B &2(p-2)B & (p-2)B \\
 B&2B &\cdots&\cdots&(p-2)B &  (p-1)B \\ 
\end{array}\label{B}
\right).
\n
\eeqa
Then, since the action of $B$ commutes 
with the multiplication of a root or a weight, 
we can write 
\beqa
\frac{\vec{\bf p}_L+\vec{\bf p}_R}{\sqrt{2}}
&=&
\sum_{k=1}^{p-1}
({\bf n}_k+{\bf j}_k B)\vec{\alpha}_k
+
{\bf r} B \,\vec{\omega}_1,\n
&\equiv&{\bf r}B\,\vec\omega_1~~~\mbox{mod $(\Lambda_R^{SU(p)})^n$}
\eeqa
as desired.
Thus we have shown that 
the Narain lattice of two-dimensional CFT compactified on a 
$(p-1)n$-dimensional torus with self-dual radius,
whose metric and the B-field are given by (\ref{G}) and (\ref{B}), respectively,  
is isomorphic to the lattice $\Gamma_{\cal C}$
(\ref{Gamma_CLambda_RLambda_W})
constructed from an error-correcting code $\cal C$ over $\FF_p$ or $\ZZ_p$ 
with the generator matrix $\left(I~B\right)$.
\\
\\
\noindent
{\it $ADE$ generalizations.}~~~As we noted previously, 
we can take $\frak g=E_6, E_7$ and $D_k$ for odd $k$ to construct 
a lattice for a code over $\FF_3$, $\FF_2$ and $\ZZ_4$, respectively, by means of 
Construction ${\rm A}_{\frak{g}}$. The lattice is given by  $\Gamma_{\cal C}^{\frak g}$ 
(\ref{Gamma_C^g}). By construction, it is almost obvious that the Narain CFT whose 
momentum-winding lattice coincides with such a lattice constructed by Construction ${\rm A}_{\frak{g}}$
is similarly obtained by a compactification on a torus with self-dual radius with 
$G_{\mu\nu}$ and $B_{\mu\nu}$ given by a tensor product of $I$ and $C_{\frak{g}}^{-1}$, and 
$B$ and $C_{\frak{g}}^{-1}$, where $C_{\frak{g}}$ is this time the Cartan matrix of the corresponding 
Lie algebra $\frak{g}$.
\\
\\
\noindent
{\it Example.}\\
(1) B-form binary codes of length $2n$. Binary codes use $\FF_2$, so $p=2$. 
The weight space of $SU(2)$ is one-dimensional, and the unique simple root $\vec\alpha_1$
is represented as $\sqrt{2}$, and the fundamental weight $\vec\omega_1$ is $\frac1{\sqrt{2}}$.
So the corresponding Narain CFT is the one compactified on an $n$-dimensional torus 
with $R=\sqrt{2}$ and $G_{\mu\nu}={\cal G}=\frac I{2}$, $B_{\mu\nu}={\cal B}=\frac B{2}$. 
Note that $G_{\mu\nu}$, $B_{\mu\nu}$ and $R$ are not independent but
there is redundancy in their values:
\beqa
G_{\mu\nu}\mapsto\lambda^{-2}G_{\mu\nu},~~~
B_{\mu\nu}\mapsto\lambda^{-2}B_{\mu\nu},~~~
R\mapsto\lambda R
\label{redundancy}
\eeqa
for $\lambda\in\RR$, which leaves the partition function invariant. 
Therefore, by taking $\lambda=\frac1{\sqrt{2}}$, we can equivalently 
say that the corresponding Narain CFT is a CFT on an $R=1$ torus with 
$G_{\mu\nu}={\cal G}=I$ and $B_{\mu\nu}={\cal B}=B$.
This is in exact agreement with the known result \cite{Dymarsky2020}.
\\
\noindent
(2) The length-two B-form ternary code. In this case $n=1$, so $B=0$. As we saw above, 
the code lattice is the same as the one made by 
Construction A${}_{\CCsub}$. The corresponding Narain CFT is obtained by 
the compactification on a two-dimensional torus with $R=\sqrt{2}$ and 
\beqa
G_{\mu\nu}&=&\frac13\left(
\begin{array}{cc}
2&1\\
1&2
\end{array}
\right), ~~~ B_{\mu\nu}=0.
\eeqa
\noindent
(3) The length-four B-form ternary codes. In this case $n=2$, so there are three choices for 
the generator matrix 
\beqa
\left(
\begin{array}{cccc}
1&0&0&0\\
0&1&0&0
\end{array}
\right),~~~~~
\left(
\begin{array}{cccc}
1&0&0&\pm1\\
0&1&\mp1&0
\end{array}
\right).
\eeqa
The first choice corresponds to the CFT compactified on a four-dimensional 
torus with $R=\sqrt{2}$ and
\beqa
G_{\mu\nu}&=&\frac13\left(
\begin{array}{cccc}
2&&1&\\
&~~2&&~~1\\
1&&2&\\
&~~1&&~~2
\end{array}
\right), ~~~ B_{\mu\nu}=0,
\eeqa
while the second and the third choices to the same $R=\sqrt{2}$ and 
\beqa
G_{\mu\nu}&=&\frac13\left(
\begin{array}{cccc}
2&&1&\\
&~~2&&~~1\\
1&&2&\\
&~~1&&~~2
\end{array}
\right), ~~~ B_{\mu\nu}=
\frac13\left(
\begin{array}{cccc}
&\pm2&&\pm1\\
\mp2&&\mp1&\\
&\pm1&&\pm2\\
\mp1&&\mp2&
\end{array}
\right).
\eeqa
%
\\
\\
\noindent
{\it 
A further extension: the $E_8$ lattice from codes over 
the Mordell-Weil (MW) group of extremal rational elliptic surfaces.}
~~~Finally, as an application of the further extension of Construction A${}_{\g}$ mentioned here, we will discuss the construction of the $E_8$  lattice from classical codes on the Mordell-Weil (MW) group of extremal rational elliptic surfaces \footnote{The $E_8$ lattice is a lattice with 
a Euclidean inner product, so it only corresponds to chiral CFTs, but 
it may be useful for constructing the $E_8$ part 
of heterotic Narain CFTs  from codes.}.

Until now, for a given code on $\FF_p$ or $\ZZ_q$, we have constructed 
a lattice by choosing one $\g$ whose $ \Lambda_W^{\fraksmall g}/\Lambda_R^{\fraksmall g}$
 is isomorphic to $\FF_p$ or $\ZZ_q$, and assigning the 
 $ \Lambda_W^{\fraksmall g}/\Lambda_R^{\fraksmall g}$
 of this $\g$ to all symbols in common.
 Here, let us consider constructing a lattice by associating 
 $ \Lambda_W^{\fraksmall g}/\Lambda_R^{\fraksmall g}$
  of a different Lie algebra for each symbol of the code.

For example, let us consider
 a ternary code generated by the generator matrix 
$
(1~1)
$
and construct a code lattice by applying the ordinary 
Construction A${}_{\CCsub}$ to the first symbol, but
Construction A${}_{\gsmall}$ with $\g=E_6$ to the 
second symbol. 
Since  $ \Lambda_W^{\fraksmall g}/\Lambda_R^{\fraksmall g}$
for both $\g=SU(3)$ and $E_6$, this kind of ``mixed" Construction A$_{\gsmall}$ is possible.
Then the associated lattice theta function is
\beqa
\sum_{c=0}^2
\Theta_{c\vec{\omega}^{SU(3)}_1,1}^{\Lambda^{SU(3)}_R}(\tau)
\Theta_{c\vec{\omega}^{E_6}_1,1}^{\Lambda^{E_6}_R}(\tau),
\label{SU(3)xE6Theta}
\eeqa
where
\beqa
\Theta_{\vec{\lambda},k}^{\Lambda}(\tau,\vec{z})&=&
\sum_{\vec{x}\in\Lambda}
q^{\frac k2 (\vec{x}+\frac{\vec{\lambda}}k)^2}
e^{2\pi i k \vec{z} (\vec{x}+\frac{\vec{\lambda}}k)}.
\eeqa
One can show that (\ref{SU(3)xE6Theta}) is modular $T$ invariant,
and also modular $S$ invariant if it is multiplied by $\eta(\tau)^8$.
Therefore, (\ref{SU(3)xE6Theta}) is the theta function 
associated with the only such rank-$8$ lattice, namely the 
$E_8$ theta function.

Similarly, one can consider a binary code generated from the same 
generator matrix, and apply Construction A to the first symbol and 
Construction A${}_{\g}$ with $\g=E_7$  to the second symbol. 
Both $SU(2)$ and $E_7$ has 
$ \Lambda_W^{\fraksmall g}/\Lambda_R^{\fraksmall g}=\ZZ_2$.
The resulting lattice theta function is 
\beqa
\sum_{c=0}^1
\Theta_{c\vec{\omega}^{SU(2)}_1,1}^{\Lambda^{SU(2)}_R}(\tau)
\Theta_{c\vec{\omega}^{E_7}_1,1}^{\Lambda^{E_7}_R}(\tau).
\label{SU(2)xE7Theta}
\eeqa
Again, this is modular invariant if it is supplemented by $\eta(\tau)^8$,
giving another expression for the $E_8$ theta function.
Clearly these correspond to orthogonal decompositions of the 
$E_8$ lattice, and similarly we can do the same for other decompositions 
related to grand unified theories such as $SU(4)\times SO(10)$ and 
$SU(5)\times SU(5)$, which are important for phenomenological applications 
of heterotic string theory and F-theory.

In fact, what's more, these rank-$8$ subgroups of $E_8$ are realized 
as singular fiber types of extremal rational elliptic surfaces, 
and the rings over which the corresponding codes are defined 
are the Mordell-Weil groups of those surfaces. 
This is not a coincidence, because the structure theorem 
of the Mordell-Weil groups of rational elliptic surfaces \cite{OguisoShioda}
guarantees that the Mordell-Weil group of a rational elliptic surface 
is a quotient of $E_8$ by the singularity of such a rational elliptic surface. 
Thus, for rational elliptic surfaces with singularity of rank-$8$, 
{\it i.e.} extremal rational elliptic surfaces, 
the MW group becomes a torsion subgroup, 
and information about the decomposition of $E_8$ 
is encoded in a code over it.
While we constructed in this paper a lattice using an isomorphism 
between the field or ring over which the code is defined and 
$ \Lambda_W^{\fraksmall g}/\Lambda_R^{\fraksmall g}$ 
of a certain Lie algebra $\g$,
by relaxing this to a homomorphism with a nontrivial kernel or co-kernel 
and constructing a code lattice, we can realize the $E_8$ lattice 
from the codes over the Mordell-Weil groups for all classified extremal rational elliptic surfaces.
%
%

A detailed account of this Letter is in preparation.

\vskip 5mm
The authors thank S.~Yata for helpful discussions, and 
A.~Dymarsky for useful comments.
The work of S.M. was supported by JSPS KAKENHI Grant Number JP23K03401,
and the work of T.O. was supported by JST SPRING, Grant Number JPMJSP2104.

%

\end{document}